\documentclass[prc,showpacs,preprintnumbers,superscriptaddress,floatfix]{revtex4}
\usepackage{dcolumn}
\usepackage{bm}
\usepackage{longtable}
\usepackage{mathrsfs}
\usepackage{graphicx,epsfig,latexsym,amssymb}
\usepackage{multirow,amsmath,array,booktabs,color}
\usepackage[section]{placeins}

\begin{document}

\title{Probing the resonance in the Dirac equation with quadruple-deformed
potentials by complex momentum representation method}
\author{Zhi Fang}
\affiliation{School of Physics and Materials Science, Anhui University, Hefei 230601,
P.R.China}
\author{Min Shi}
\affiliation{School of Physics and Materials Science, Anhui University, Hefei 230601,
P.R.China}
\affiliation{RIKEN Nishina Center, Wako 351-0198, Japan}
\author{Jian-You Guo}
\email[E-mail:]{jianyou@ahu.edu.cn}
\affiliation{School of Physics and Materials Science, Anhui University, Hefei 230601,
P.R.China}
\author{Zhong-Ming Niu}
\affiliation{School of Physics and Materials Science, Anhui University, Hefei 230601,
P.R.China}
\affiliation{Interdisciplinary Theoretical Science Research Group, RIKEN, Wako 351-0198, Japan}
\author{Haozhao Liang}
\affiliation{RIKEN Nishina Center, Wako 351-0198, Japan}
\affiliation{Interdisciplinary Theoretical Science Research Group, RIKEN, Wako 351-0198, Japan}
\affiliation{Department of Physics, Graduate School of Science, The University of Tokyo,
Tokyo 113-0033, Japan}
\author{Shi-Sheng Zhang}
\affiliation{School of Physics and Nuclear Energy Engineering, Beihang University,
Beijing 100191, China}
\date{\today }

\begin{abstract}
Resonance plays critical roles in the formation of many physical phenomena,
and many techniques have been developed for the exploration of resonance. In
a recent letter [Phys. Rev. Lett. 117, 062502 (2016)], we proposed a
new method for probing single-particle resonances by solving the Dirac equation in complex momentum
representation for spherical nuclei. Here, we extend this method to deformed
nuclei with theoretical formalism presented. We elaborate numerical details,
and calculate the bound and resonant states in $^{37}$Mg. The results are
compared with those from the coordinate representation calculations with a
satisfactory agreement. In particular, the present method can expose clearly
the resonant states in complex momentum plane and determine precisely the
resonance parameters for not only narrow resonances but also broad resonances
that were difficult to obtain before.
\end{abstract}

\pacs{21.60.Jz,21.10.Pc,25.70.Ef}
\maketitle




\section{Introduction}

Resonance is one of the most striking phenomena in the whole range of scattering
experiments and appears widely in atomic, molecular, and nuclear physics~%
\cite{Taylor72}. Resonance plays critical roles in the formation of many
physical phenomena such as the quantum halos~\cite{Jensen04}. Theoretical
explanation of halo in $^{11}$Li~\cite{Meng96}, prediction of giant halo in
Zr and Ca isotopes~\cite{Meng98,Meng02}, and understanding of deformed halo
in $^{31}$Ne and $^{44}$Mg~\cite{Hamamoto10,Zhou10} are mainly attributed to
consider the contributions from the continuum, especially the resonances in
the continuum. The change of traditional magic numbers in these nuclei with
unusual neutron-to-proton ratio can be understood in terms of the shell
structure of resonant levels~\cite{Hamamoto12}. It is also found that the contribution of the
continuum to the giant resonances mainly comes from single-particle
resonances~\cite{Curutchet89,Cao02}. Resonance is also closely relevant to
the nucleosynthesis of chemical elements in the Universe~\cite%
{SZhang12,Faestermann15}. Therefore, research on resonance is one
of the hottest topics in different branches of physics.

So far, a series of methods have been proposed for resonance,
including the scattering phase shift method, the analytic continuation
in the coupling constant (ACCC) approach, \textit{R}-matrix method, \textit{S%
}-matrix method, Green's function method, etc. These methods have gained
success in handling unbound problems. Even so, one hopes to establish a
unified theory, which can deal with both bound states and resonant states on
the same footing. The complex scaled method (CSM) introduced in Refs.~\cite%
{ABC71} would satisfy this requirement.

In the CSM, the wave functions adopted for the resonant states are
square-integrable, and thus it is not necessary to use the asymptotic boundary
conditions. Moreover, the complex scaled equation can be solved by using the
bound-state methods, in which the bound states and resonant states are
processed equally. These advantages enable the application of CSM to
different theoretical frameworks, including the combinations with the
few-body models~\cite{Arai06}, shell models~\cite{Michel08,Michel09}, and
Hartree-Fock theories~\cite{Kruppa97,Kruppa14}. More applications can be found
in Refs.~\cite{Moiseyev98,Myo14,Carbonell14}. Recently, we have applied the
CSM to explore the resonances in spherical nuclei~\cite%
{GuoPRC10,GuoCPC10,Zhu14,MShi15} and deformed nuclei~\cite%
{liu12,MShi14,XShi16} in a satisfactory agreement with those obtained by other
methods.

Although it can describe the bound states and resonant
states in a unified way, the CSM still has some shortcomings. For example, in order to determine accurately the
resonance parameters, repeated diagonalization of the Hamiltonian is
required in the complex scaling calculations. In addition, the CSM is only applicable to
the dilation analytic potentials. For the systems like nuclei, the mean-field potentials
for nucleon movement are similar to the Woods-Saxon potentials. There appear
singularities when the complex rotation angle $\theta=\tan^{-1}\left(\pi
a/R\right)$. Hence, the CSM is only effective in the interval of $0<\theta<\tan^{-1}\left(\pi a/R\right) $ for the resonances in nuclei, which confines
the application of the CSM for very broad resonances, while the broad resonances
deserve more attentions for their roles in exotic phenomena.

In order to hold the advantages of CSM and avoid its shortcomings,
the complex momentum representation (CMR) method has been proposed.
In Ref.~\cite{Berggren68}, the Schr\"{o}dinger
equation was formalized using momentum representation. This method has
avoided all the defects in the CSM, and has been used to explore the bound
states~\cite{Sukumar79,Kwon78} and resonant states~\cite{Hagen06,Deltuva15}
in the nonrelativistic case, and used as the so-called ``Berggren representation'' in the shell-model calculations~\cite{Liotta96,Michel02}. Considering that the
relativistic resonances are widely concerned, almost all the methods for
resonances have been extended to the relativistic framework~\cite%
{Horodecki00,Fuda01,SZhang04,LZhang08,Grineviciute12,Lu12,TSun14}, including
the relativistic CSM~\cite{GuoPRC10,MShi14} and relativistic complex scaled
Green's function method~\cite{MShi15}. Recently, we applied the
CMR method to the relativistic mean-field (RMF) framework and
established the RMF-CMR method for the resonances in the spherical case~\cite{Li16},
in which both bound states and resonant states have been treated on the same footing.
The RMF-CMR method gathers the advantages of the RMF and CMR, and is able to describe self-consistently nuclear bound states and resonant states in the relativistic framework. Due to these advantages, to extend the method to deformed nuclei is worthwhile as most of nuclei with deformation.

In this paper we develop the relativistic version of CMR method for deformed
nuclei, in which the Dirac equation describing deformed nuclei is processed into a set of coupled differential equations by the coupled-channel method and the set of coupled differential equations are solved using the complex momentum representation technique. We will first present the theoretical formalism, and then elaborate numerical details. Taking the nucleus $^{37}$Mg as an example, we calculate the bound and resonant states, and compare with those obtained in coordinate representation with the ACCC method.

\section{Formalism}

Considering that the relativistic mean-field theory is very successful in
describing various nuclear phenomena~\cite%
{Serot86,Ring96,Vretenar05,Meng06,Niksic11,Liang15,Meng15} and nuclear inputs in astrophysics~\cite{Sun08,Niu09,Xu13,Niu13,Niu13b}, we explore the single-particle resonances in
deformed nuclei based on the relativistic mean-field theory with the Dirac equation as
\begin{equation}
\left[ \vec{\alpha}\cdot \vec{p}+\beta \left( M+S\right) +V\right] \psi
=\varepsilon \psi,  \label{Diraceq}
\end{equation}%
where $M$ ($\vec{p}$) is the nucleon mass (momentum), $\vec{\alpha}$ and $%
\beta $ are the Dirac matrices, and $S$ and $V$ are the scalar and vector
potentials, respectively. The details of the RMF theory can refer to the
literatures~\cite{Serot86,Ring96,Vretenar05,Meng06}.

The solutions of Eq.~(\ref{Diraceq}) include the bound states, resonant
states, and nonresonant continuum. The bound states can be obtained with
conventional methods. For the resonant states, many techniques have been
developed, while some of them exist certain shortcomings. In Ref.~\cite{Li16}, we
proposed a new method by solving the Dirac equation in
complex momentum representation for spherical nuclei. In the present work, we extend this
method to deformed nuclei. Without loss of generality, only the axially
symmetrical quadruple deformation is considered here, $V(\vec{r})$ and $S(\vec{r})$ are taken as
\begin{equation}
\left\{
\begin{array}{lcl}
V\left( \vec{r}\right) & = & V_{0}f(r)-\beta _{2}V_{0}k\left( r\right)
Y_{20}\left( \vartheta ,\varphi \right) , \\
S\left( \vec{r}\right) & = & S_{0}f(r)-\beta _{2}S_{0}k\left( r\right)
Y_{20}\left( \vartheta ,\varphi \right) ,%
\end{array}%
\right.
\end{equation}%
where $\beta _{2}$ is the quadruple deformation parameter. Similar to Ref.~%
\cite{Li10}, a Woods-Saxon type potential is adopted with $f\left( r\right) =%
\frac{1}{1+\exp \left[ \left( r-R\right) /a\right] }$ and $k\left( r\right) =%
\frac{rdf\left( r\right) }{dr}$. In order to explore the resonances in
deformed nuclei, we transform Eq.~(\ref{Diraceq}) into momentum
representation as
\begin{equation}
\int d\vec{k}^{\prime }\left\langle \vec{k}\right\vert H\left\vert \vec{k}%
^{\prime }\right\rangle \psi \left( \vec{k}^{\prime }\right) =\varepsilon
\psi \left( \vec{k}\right) ,  \label{MomentDireq}
\end{equation}%
where $H=\vec{\alpha}\cdot \vec{p}+\beta \left( M+S\right) +V$ is the Dirac
Hamiltonian, $\psi \left( \vec{k}\right) $ is the corresponding wave
function in momentum representation, and $\left\vert \vec{k}\right\rangle $
represents the wave function of a free particle with wave vector $\vec{k}=$ $%
\vec{p}/\hbar $. In order to solve the Dirac equation (\ref{MomentDireq})
for deformed system, the coupled-channel method is adopted, the wave
function is expanded as
\begin{equation}
\psi \left( \vec{k}\right) =\psi _{m_{j}}\left( \vec{k}\right)
=\sum\limits_{lj}\left(
\begin{array}{c}
f^{lj}(k)\phi _{ljm_{j}}\left( \Omega _{k}\right) \\
g^{lj}(k)\phi _{\tilde{l}jm_{j}}\left( \Omega _{k}\right)%
\end{array}%
\right) ,\left( \tilde{l}=2j-l\right)  \label{wavefunctions}
\end{equation}%
with
\begin{equation*}
\phi _{ljm_{j}}\left( \Omega _{k}\right) =\sum\limits_{m_{s}}\langle lm\frac{%
1}{2}m_{s}|jm_{j}\rangle Y_{lm}\left( \Omega _{k}\right) \chi _{m_{s}},
\end{equation*}%
where $f^{lj}(k)$ and $g^{lj}(k)$ are the radial components of Dirac spinors
in momentum representation, $l$ and $m$ are the quantum numbers of the
orbital angular momentum and its projection on the third axis, $j$ and $%
m_{j} $ are the quantum numbers of the total angular momentum and its
projection on the third axis, and $\chi _{m_{s}}$ is the spin wave function
with the third component of spin angular momentum $m_{s}$. It should be
emphasized that the projection of the total angular momentum on the third
axis $m_{j}$ and the parity $\pi $ are good quantum numbers for an axially
deformed system.

Putting the wave function (\ref{wavefunctions}) into the equation (\ref%
{MomentDireq}), the Dirac equation becomes
\begin{equation}
\left\{
\begin{array}{c}
Mf^{lj}(k)-kg^{lj}(k)+\sum\limits_{l^{\prime }j^{\prime }}\int k^{\prime
2}dk^{\prime }V^{+}(l^{\prime },j^{\prime },p,q,l,j,m_{j},k,k^{\prime
})f^{l^{\prime }j^{\prime }}(k^{\prime })=\varepsilon f^{lj}\left( k\right) ,
\\
-kf^{lj}(k)-Mg^{lj}(k)+\sum\limits_{l^{\prime }j^{\prime }}\int k^{\prime
2}dk^{\prime }V^{-}(\tilde{l}^{\prime },j^{\prime },p,q,\tilde{l}%
,j,m_{j},k,k^{\prime })g^{l^{\prime }j^{\prime }}(k^{\prime })=\varepsilon
g^{lj}\left( k\right) ,%
\end{array}%
\right.   \label{integraleq}
\end{equation}%
with
\begin{eqnarray}
&&V^{+}(l^{\prime },j^{\prime },p,q,l,j,m_{j},k,k^{\prime })  \notag \\
&=&\left( -\right) ^{l}i^{l+l^{\prime }}\frac{2}{\pi }\int r^{2}dr\left[
V\left( r\right) +S\left( r\right) \right] j_{l}\left( kr\right)
j_{l^{\prime }}\left( k^{\prime }r\right) \sum\limits_{m_{s}}\left\langle
lm\right\vert Y_{pq}\left( \Omega _{r}\right) \left\vert l^{\prime
}m^{\prime }\right\rangle \langle lm\frac{1}{2}m_{s}|jm_{j}\rangle \langle
l^{\prime }m^{\prime }\frac{1}{2}m_{s}|j^{\prime }m_{j}\rangle ,  \label{vps}
\\
&&V^{-}(\tilde{l}^{\prime },j^{\prime },p,q,\tilde{l},j,m_{j},k,k^{\prime })
\notag \\
&=&\left( -\right) ^{\tilde{l}}i^{\tilde{l}+\tilde{l}^{\prime }}\frac{2}{\pi
}\int r^{2}dr\left[ V\left( r\right) -S\left( r\right) \right] j_{\tilde{l}%
}\left( kr\right) j_{\tilde{l}^{\prime }}\left( k^{\prime }r\right)
\sum\limits_{m_{s}}\left\langle \tilde{l}\tilde{m}\right\vert Y_{pq}\left(
\Omega _{r}\right) \left\vert \tilde{l}^{\prime }\tilde{m}^{\prime
}\right\rangle \langle \tilde{l}\tilde{m}\frac{1}{2}m_{s}|jm_{j}\rangle
\langle \tilde{l}^{\prime }\tilde{m}^{\prime }\frac{1}{2}m_{s}|j^{\prime
}m_{j}\rangle ,  \label{vms}
\end{eqnarray}%
where $j_{l}\left( kr\right)$ $[j_{\tilde{l}}\left( kr\right) ]$ are the
spherical Bessel functions of order $l$ $[\tilde{l}]$. Equation~(\ref{integraleq})
is a set of coupled integral equations. Its solution is difficult to obtain by the
conventional methods especially for the resonant states. By turning the
momentum integral into a sum over a finite set of
points $k$ and $dk$ with a set of weights $w$, the integral equation~(\ref{integraleq}) becomes a matrix equation
\begin{equation}
\left\{
\begin{array}{c}
Mf^{lj}(k_{a})-k_{a}g^{lj}(k_{a})+\sum\limits_{l^{\prime }j^{\prime
}}\sum\limits_{b}w_{b}k_{b}^{2}V^{+}(l^{\prime },j^{\prime
},p,q,l,j,m_{j},k_{a},k_{b})f^{l^{\prime }j^{\prime }}(k_{b})=\varepsilon
f^{lj}\left( k_{a}\right) , \\
-k_{a}f^{lj}(k_{a})-Mg^{lj}(k_{a})+\sum\limits_{l^{\prime }j^{\prime
}}\sum\limits_{b}w_{b}k_{b}^{2}V^{-}(\tilde{l}^{\prime },j^{\prime },p,q,%
\tilde{l},j,m_{j},k_{a},k_{b})g^{l^{\prime }j^{\prime }}(k_{b})=\varepsilon
g^{lj}\left( k_{a}\right) .%
\end{array}%
\right.   \label{matrixeq}
\end{equation}%
In Eq.~(\ref{matrixeq}), the Hamiltonian matrix is not symmetric. For
simplicity in computation, we symmetrize it by the following transformation
\begin{equation}
\left\{
\begin{array}{c}
\mathbf{f}(k_{a})=\sqrt{w_{a}}k_{a}f(k_{a}), \\
\mathbf{g}(k_{a})=\sqrt{w_{a}}k_{a}g(k_{a}),%
\end{array}%
\right.
\end{equation}%
which gives us a symmetric matrix in the momentum representation as
\begin{equation}
\left\{
\begin{array}{c}
\sum\limits_{b}\left[ M\delta _{ab}\mathbf{f}^{lj}(k_{b})+\sum\limits_{l^{%
\prime }j^{\prime }}\sqrt{w_{a}w_{b}}k_{a}k_{b}V^{+}(l^{\prime },j^{\prime
},p,q,l,j,m_{j},k_{a},k_{b})\mathbf{f}^{l^{\prime }j^{\prime
}}(k_{b})-k_{a}\delta _{ab}\mathbf{g}^{lj}(k_{b})\right] =\varepsilon
\mathbf{f}^{lj}\left( k_{a}\right) , \\
\sum\limits_{b}\left[ -k_{a}\delta _{ab}\mathbf{f}^{lj}(k_{b})-M\delta _{ab}%
\mathbf{g}^{lj}(k_{b})+\sum\limits_{l^{\prime }j^{\prime }}\sqrt{w_{a}w_{b}}%
k_{a}k_{b}V^{-}(\tilde{l}^{\prime },j^{\prime },p,q,\tilde{l}%
,j,m_{j},k_{a},k_{b})\mathbf{g}^{l^{\prime }j^{\prime }}(k_{b})\right]
=\varepsilon \mathbf{g}^{lj}\left( k_{a}\right) .%
\end{array}%
\right.   \label{symmetrymatrix}
\end{equation}%
So far, to solve the Dirac equation (\ref{Diraceq}) becomes an eigensolution
problem of the symmetric matrix. All the bound and resonant states can be
obtained simultaneously by diagonalizing the Hamiltonian in Eq.~(\ref%
{symmetrymatrix}). Compared with Refs.~\cite{Li10,Xu15}, where the bound
solutions are obtained by solving a set of coupled differential equations
and every resonant state is handled solely by the scattering phase
shift method or ACCC approach, there is no doubt that the present method is more convenient.

The diagonalization of the Hamiltonian matrix in Eq.~(\ref{symmetrymatrix}) can provide us the energies and wave functions in the momentum representation. If we regard the wave functions in the coordinate representation, the following transformation is introduced
\begin{equation}
\psi \left( \vec{r}\right) =\left\langle \vec{r}\right\vert \psi \rangle =%
\frac{1}{\left( 2\pi \right) ^{3/2}}\int d\vec{k}e^{i\vec{k}\cdot \vec{r}%
}\psi \left( \vec{k}\right) .  \label{Rwavefunctions}
\end{equation}%
For an axially deformed nucleus, putting the wavefunctions (\ref%
{wavefunctions}) into the equation (\ref{Rwavefunctions}), we obtain the
Dirac spinors in the coordinate space as
\begin{equation}
\psi \left( \vec{r}\right) =\psi _{m_{j}}\left( \vec{r}\right)
=\sum\limits_{lj}\left(
\begin{array}{c}
f^{lj}(r)\phi _{ljm_{j}}\left( \Omega _{r}\right) \\
g^{lj}(r)\phi _{\tilde{l}jm_{j}}\left( \Omega _{r}\right)%
\end{array}%
\right) ,\label{radialwavef}
\end{equation}%
with the radial components
\begin{equation*}
\left\{
\begin{array}{c}
f^{lj}(r)=i^{l}\sqrt{\frac{2}{\pi }}\sum\limits_{a}\sqrt{w_{a}}%
k_{a}j_{l}\left( k_{a}r\right) \mathbf{f}^{lj}(k_{a}), \\
g^{lj}(r)=i^{\tilde{l}}\sqrt{\frac{2}{\pi }}\sum\limits_{a}\sqrt{w_{a}}%
k_{a}j_{\tilde{l}}\left( k_{a}r\right) \mathbf{g}^{lj}(k_{a}).%
\end{array}%
\right.
\end{equation*}

\section{Numerical details and results}

Based on the preceding formalism, we explore the resonances in real nuclei.
Before starting these calculations, we clarify several key points: (i) the coupled-channel method is adopted in solving the Dirac equation for deformed nuclei, where the wave functions are expanded with different channels labelled as $lj$. The sum over $lj$ in Eq.~(\ref{wavefunctions}) needs to be restricted to a limited number $N_{c}$; (ii) the momentum integral in Eq.~(\ref{integraleq}) is from zero to infinity, it needs to be truncated into a large enough momentum $k_{\text{max}}$. When $k_{\text{max}} $ is fixed, the integral can be calculated by a sum shown in Eq.~(\ref%
{symmetrymatrix}). As a sum with evenly spaced $dk$ and a constant weight $%
w_{a}$ converges slowly, here it is replaced by the
Gauss-Legendre quadrature with a finite grid number $N_{l}$. Through these processions, the Hamiltonian in Eq.~(\ref{symmetrymatrix}) becomes a $2N_{c}N_{l}\times 2N_{c}N_{l}$ matrix. In the actual calculations, the momentum is truncated to $k_{\text{max}}=4.0$~fm$^{-1}$, which is sufficient for all the concerned resonances. The grid number of the Gauss-Legendre quadrature $N_l=120$ is used for the momentum
integral along the contour, which is enough to ensure the convergence with respect to numbers of discretization points. The coupled-channel number $N_c=8$ is taken for the wave function expansions, which is enough to ensure the required precision.

With these parameters designed, we explore the resonances in deformed nuclei with $^{37}$Mg as an example. For the comparison with the ACCC calculations, the
parameters in the scalar potential $S$ and vector potential $V$ adopted are
the same as those in Ref.~\cite{Xu15}. The Dirac equation is solved by
diagonalizing the $2N_{c}N_{l}\times 2N_{c}N_{l}$ matrix in Eq.~(\ref{symmetrymatrix}) along an appropriate contour of momentum integral. The contour is required to be large enough to
expose all the concerned resonances.

\begin{figure}[tbp]
\includegraphics[width=8.5cm]{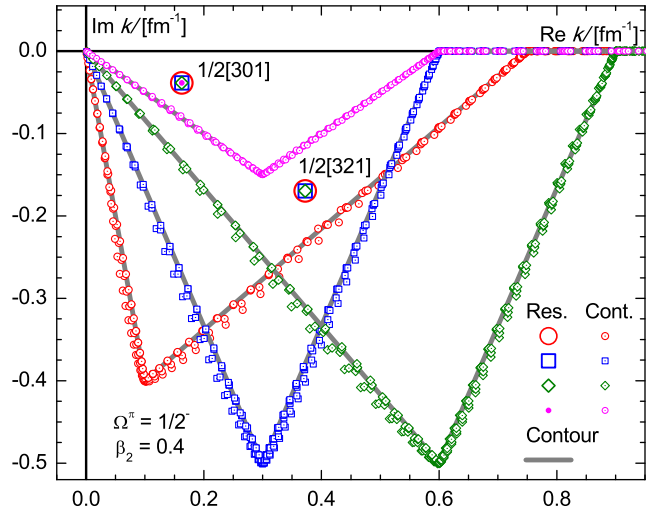}
\caption{(Color online) Single-particle spectra in $^{37}$Mg for the states $\Omega^\pi=1/2^-$ with $\beta_2=0.4$ in the complex $k$ plane with four different contours. The open red circles, blue squares, olive diamonds, and magenta dots represent the resonances obtained in four different contours, respectively. The smaller labels with a dot inside represent the continuum. The gray lines represent the corresponding contours of momentum integral.}
\label{Fig1}
\end{figure}

To single out a large enough contour, we first check the dependence of the calculations on the contour. For this purpose, we have explored the resonant states in $^{37}$Mg with four different contours, which are displayed in Fig.~\ref{Fig1} for the states $\Omega^{\pi}=1/2^-$ with $\beta_2=0.4$. Similar to the spherical case~\cite{Li16}, the resonant states have nothing to do with the choice of the contour. With the change of integral contour, the continuous spectra follow the contour, while the resonant states always stay at their original positions. When the contour becomes deeper from the magenta color to the blue color, the continuous spectra drop down with the contour, the resonant state $1/2[301]$ does not move. Similarly, when the contour moves from left (red color) to right (olive color) or from right to left, the continuous spectra follow the contour, while the resonant states $1/2[301]$ and $1/2[321]$ remain their own positions. These indicate that the physical resonant states obtained by the present method are indeed independent on the contour. Hence, we can select a large enough contour to expose all the concerned resonances. This conclusion is in agreement with that in the spherical case~\cite{Li16}.

\begin{figure}[tbp]
\includegraphics[width=8.5cm]{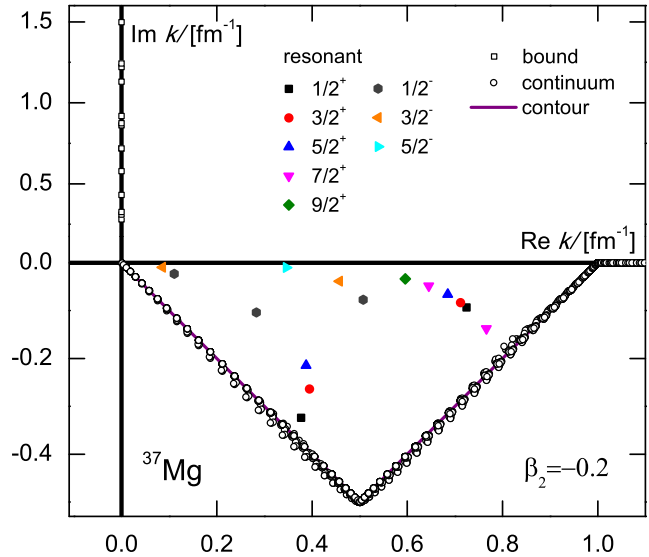}
\caption{(Color online) Single-particle spectra in $^{37}$Mg for the states $\Omega ^{\pi }=%
\frac{1}{2}^{\pm },\frac{3}{2}^{\pm },\cdots ,\frac{9}{2}^{\pm }$ with $\beta_2=-0.2$ in
the complex momentum plane. The bound states, resonant states with different quantum numbers, and continuum are marked with different labels, while the purple solid line represents the contour of momentum integral in complex momentum plane.}
\label{Fig2}
\end{figure}

Using the triangle contour with the four points $k=0$~fm$^{-1}$, $k=0.5-i0.5$~fm$^{-1}$, $k=1.0$~fm$^{-1}$, and $k_{%
\text{max}}=4.0$~fm$^{-1}$, all the concerned bound and resonant states in $%
^{37}$Mg can be obtained over the range of deformation. An illustrated
result is displayed in Fig.~\ref{Fig2} for the states $\Omega ^{\pi }=%
\frac{1}{2}^{\pm },\frac{3}{2}^{\pm },\cdots ,\frac{9}{2}^{\pm }$ with $\beta _{2}=-0.2$. From Fig.~\ref{Fig2}, it can be seen that the bound states are exposed clearly on the imaginary axis, the resonant states are isolated from the continuum in the
fourth quadrant, and the continuous spectra follow the integral contour. In the
region of resonant states, there are fifteen resonant states exposed in the
present calculations. Some resonant states are close to the real $k$ axis,
which correspond to the narrow resonances with smaller width. Some other
resonant states are far away from the real $k$ axis, which are broad
resonances. In other words, the current calculations have provided us not only the narrow resonances but also the broad resonances as long as the momentum contour covers the range of resonances.

\begin{figure}[tbp]
\includegraphics[width=8.5cm]{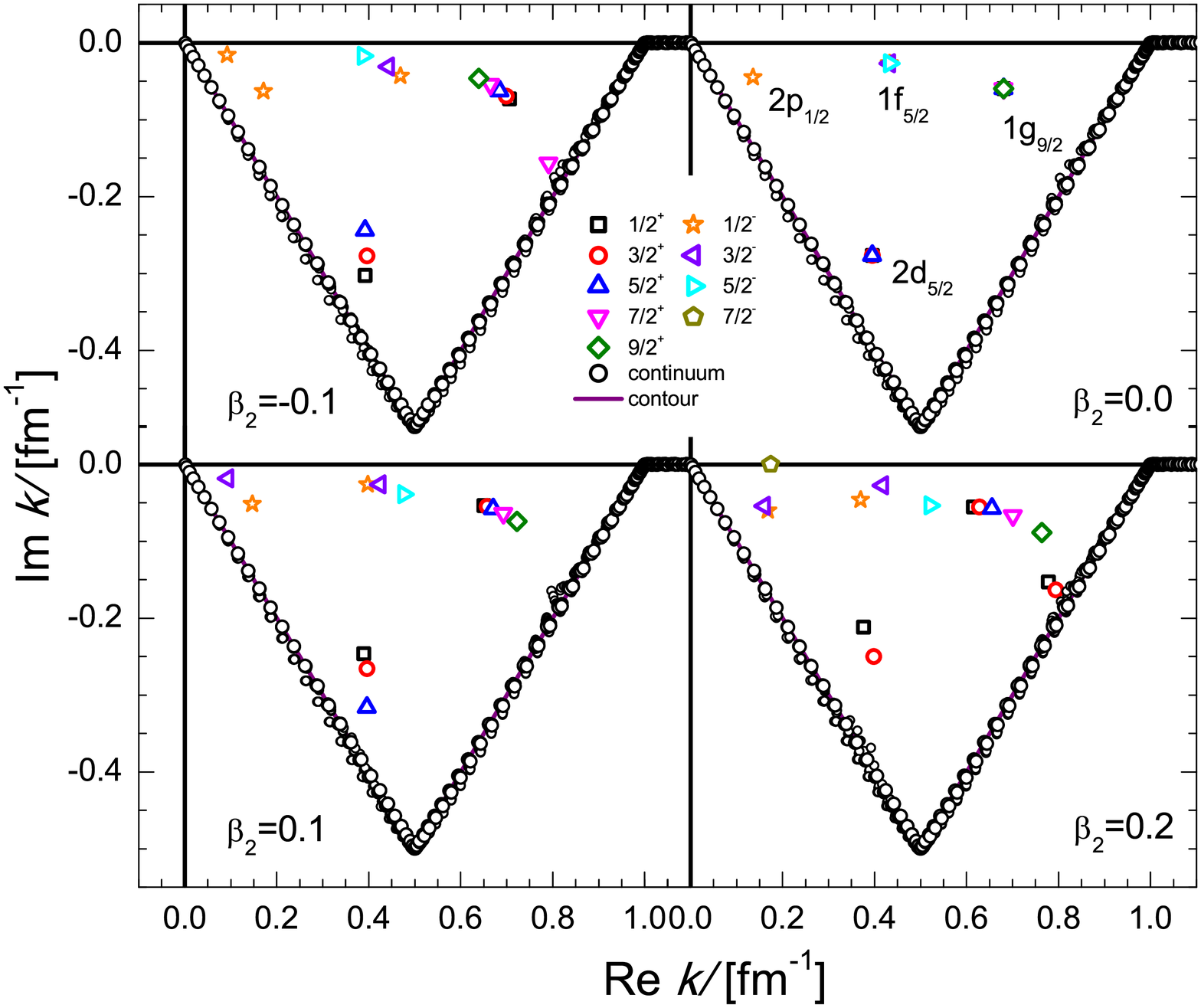}
\caption{(Color online) Single-particle resonances for the states $\Omega ^{%
\protect\pi }=\frac{1}{2}^{\pm},\frac{3}{2}^{\pm},\cdots ,\frac{9}{2}^{\pm}$ in
the complex momentum plane with several different deformations.
The outstanding labels denote the resonant states with the
corresponding quantum numbers, while the black open circles and purple solid line represent
the continuum and the integral contour in the complex momentum plane,
respectively.}
\label{Fig3}
\end{figure}

As we focus on the resonances in the deformed nuclei, it is interesting to observe intuitively the dependence of resonances on deformation. In Fig.~\ref{Fig3}, we show the $\Omega ^{\pi }=%
\frac{1}{2}^{\pm},\frac{3}{2}^{\pm},\cdots ,\frac{9}{2}^{\pm}$ resonant states with several different deformations. When $\beta _{2}=0$, the system is spherically symmetric, there appear four resonant states $2p_{1/2}$, $2d_{5/2}$, $1f_{5/2}$, and $1g_{9/2}$ in the complex momentum plane. When the spherical symmetry is broken, the degenerate states $2d_{5/2}$ and $1f_{5/2}$ respectively split into three resonant states and the degenerate state $1g_{9/2}$ into five resonant states. Their positions in the complex momentum plane depend on the deformation. With the development of deformation, some resonant states disappear and some other resonant states appear in the current region of momentum, which can be found in
Fig.~\ref{Fig3} with $\beta _{2}=0.2$.

\begin{figure}[tbp]
\includegraphics[width=8.5cm]{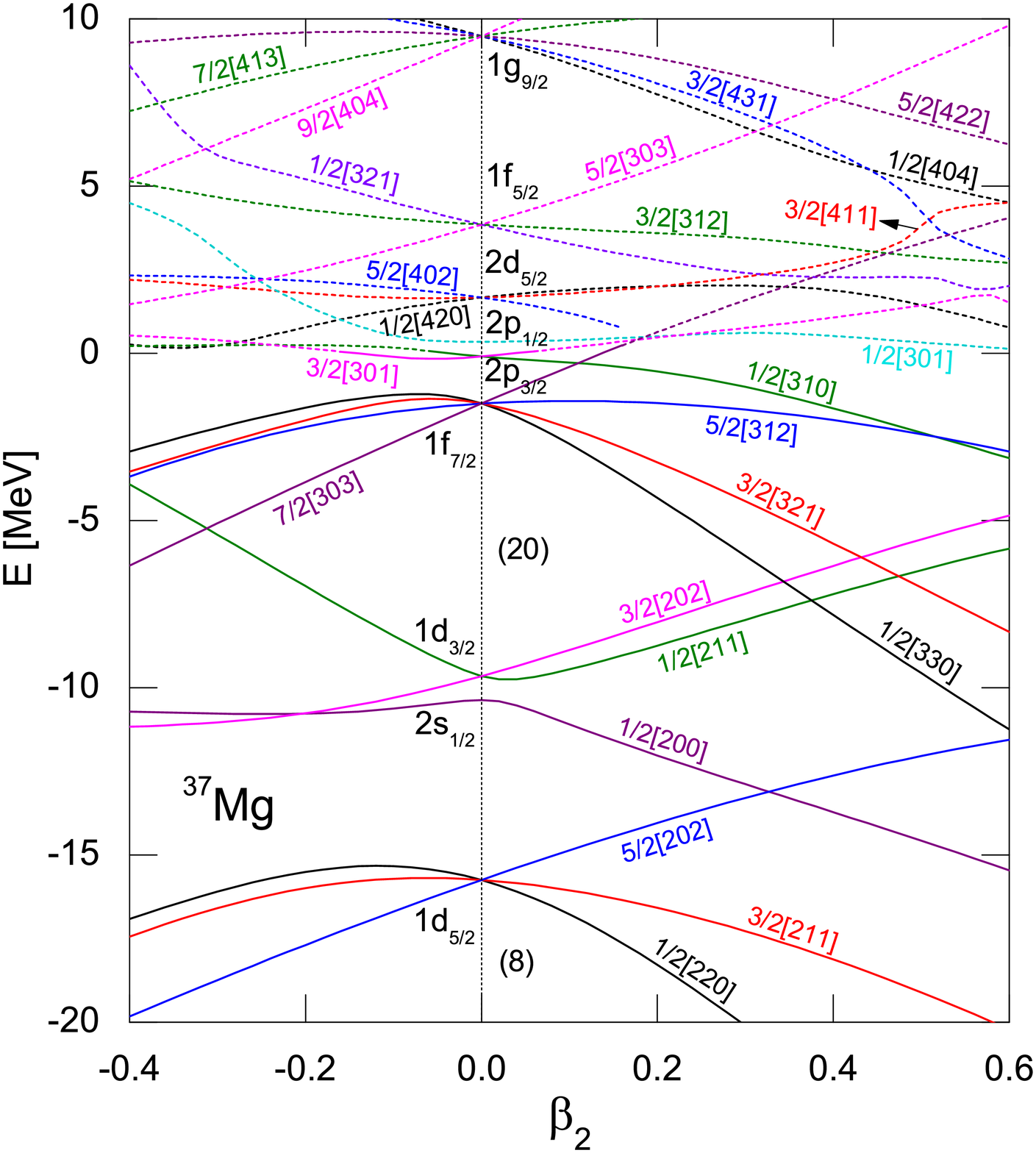}
\caption{(Color online) The evolution of single-particle energies to
deformation for all the concerned bound and resonant states, where the bound
states are marked by the solid lines and the resonant states by the dashed
lined with the Nilsson labels on the lines and the corresponding spherical labels in the position $\beta _{2}=0$.}
\label{Fig4}
\end{figure}

\begin{figure}[tbp]
\includegraphics[width=8.5cm]{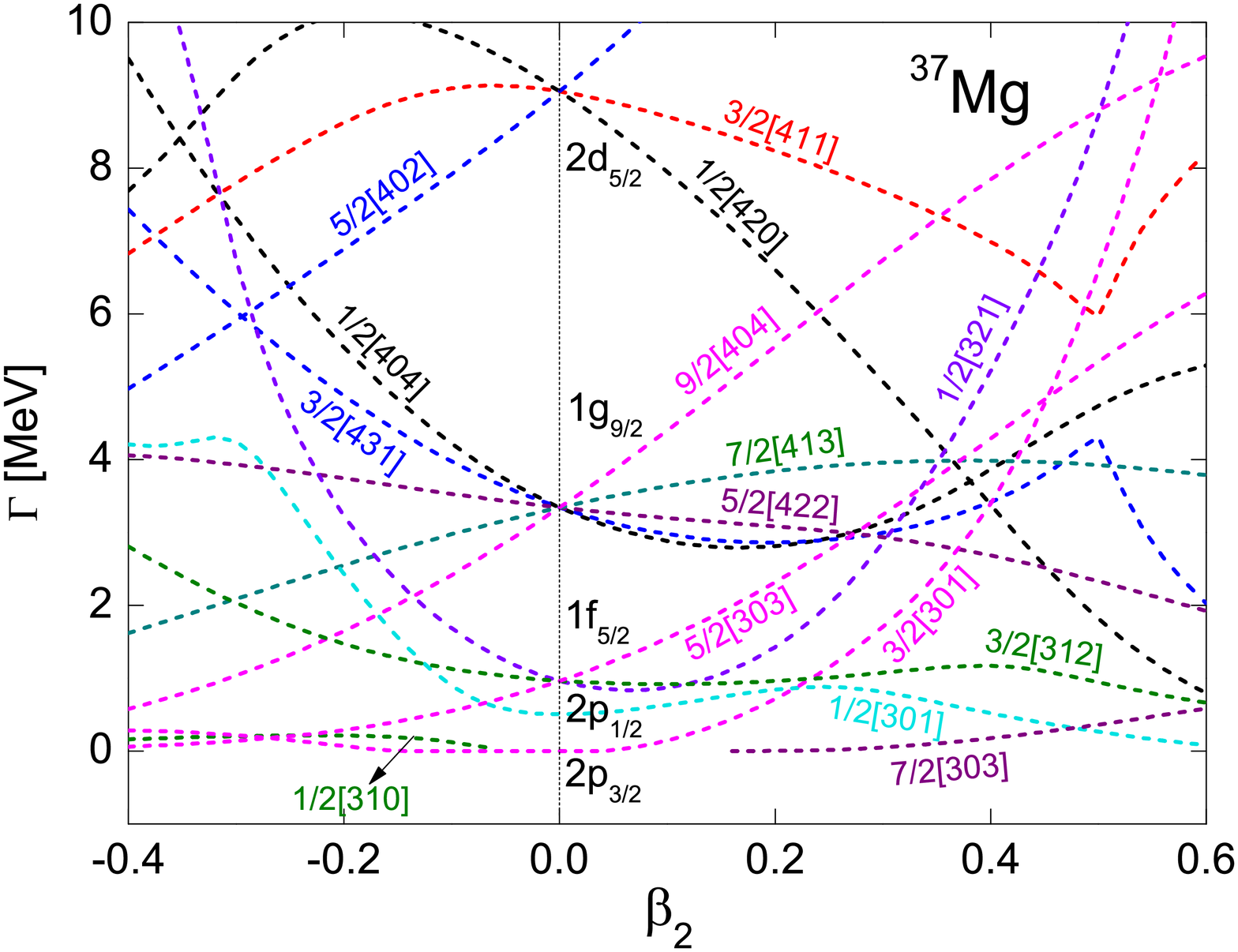}
\caption{(Color online) The evolution of widths to deformation for all
the concerned resonant states. They are marked by the
color dashed lines with the Nilsson labels
on the lines and the corresponding spherical labels in the position $%
\beta _{2}=0$.}
\label{Fig5}
\end{figure}

From Figs.~\ref{Fig2} and \ref{Fig3}, it is seen that the bound states and resonant states together with their evolutions to deformation can be obtained in the present calculations. For the resonant states, we can read the real and imaginary parts of their wavevectors from the complex $k$ plane, and then we can calculate the resonance parameters like energy and width in terms of $E_{r}+iE_{i}=E_{r}-i\Gamma /2=\sqrt{k^{2}+M^{2}}-M$. The calculated single-particle energies varying with deformation is shown in Fig.~\ref{Fig4} for the bound states and resonant states, where the bound levels are marked
by the solid line and the resonant levels by the dashed lines with the Nilsson labels
on the lines and the corresponding spherical labels in the position $\beta_{2}=0$.
For the resonant states, we show the single-particle energies in Fig.~\ref{Fig4} together with the corresponding widths in Fig.~\ref{Fig5}.

In comparison with the coupled-channel calculations in the coordinate
representation, on one hand, it is found that all the available bound levels and their evolutions to $%
\beta_{2}$ are the fully same as those shown in Ref.~\cite{Xu15}. For the resonant levels, our results are also in agreement with those obtained by the ACCC in Ref.~\cite{Xu15}.
On the other hand, in addition to the levels with the spherical labels $1f_{7/2}$, $2p_{3/2}$, $2p_{1/2}$, and $1f_{5/2}$, we also have obtained the resonant levels with the spherical labels $2d_{5/2}$ and $1g_{9/2}$. Especially for
the states with the spherical label $2d_{5/2}$ which locate at the middle of $2p_{1/2}$ and $1f_{5/2}$, the calculations in Ref.~\cite{Xu15} have not given out the results on these states.
Furthermore, a notable phenomenon appears in the resonant level $5/2[402]$. With the increasing of deformation in the probate side, its energy drops down while its width goes up rapidly.
Eventually, the resonant state disappears at the large deformation, since it
is difficult for particle to populate on the level with too short life time.

The evolution of the widths to deformation for the concerned resonant
states is shown in Fig.~\ref{Fig5}. Similar to the energies, there exists the shell structure in
these widths. Especially for the spherical case, the gap appearing in the
widths between $2d_{5/2}$ and $1g_{9/2}$ is very large. Compared with the
energy, the order is different for the width. Namely, although the energies
of the $2d_{5/2}$ states are lower, the corresponding widths are
larger due to the smaller centrifugal barrier.

\begin{figure}[tbp]
\includegraphics[width=8.5cm]{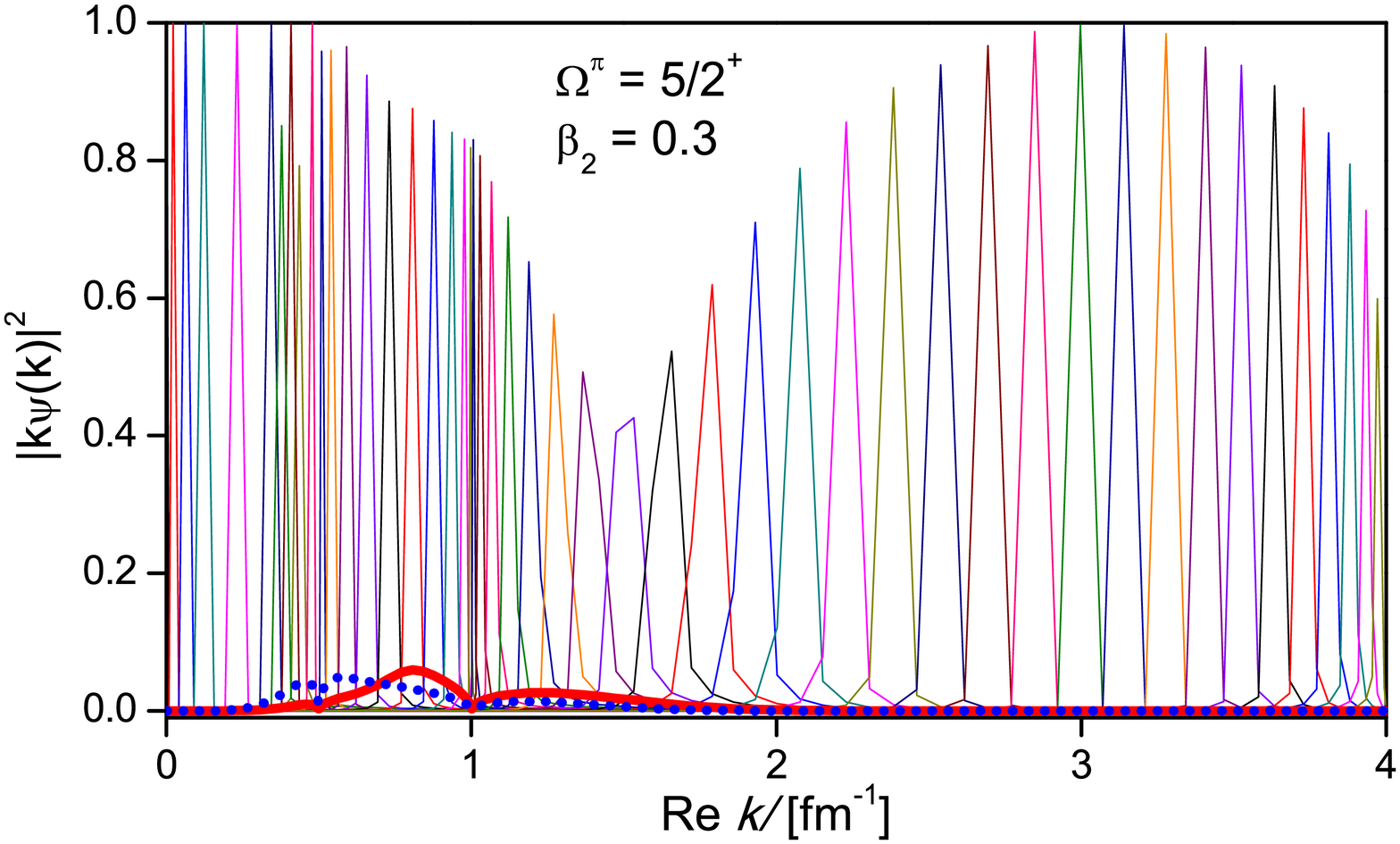}
\caption{(Color online) Radial-momentum probability distributions for the
states $\Omega^\protect\pi=5/2^+$ with $\protect\beta_2=0.3$, where the blue dashed line and red solid line correspond respectively to the bound state and resonant state, and the
others correspond to the background of the continuum.}
\label{Fig6}
\end{figure}

In addition to the energy spectra, we have also obtained the wave functions for
deformed nuclei in the momentum space. In Fig.~\ref{Fig6}, we show
the radial-momentum probability distributions (RMPD) for the single-particle
states $\Omega^\pi=5/2^+$ with the quadruple deformation $\beta_2=0.3$.
It can be seen that there are two single-particle states with their RMPD expanded much wider
than the surrounding states. By checking their energies, it is found that
the blue dashed line corresponds to the bound state $5/2[202]$ and the red solid line is
the resonant state $5/2[422]$. The other states, their RMPD display sharp peaks at different
values of $k$, corresponding to the free particles. These results agree with the
Heisenberg uncertainty principle: a less well-defined momentum corresponds
to a more well-defined position for the bound and resonant states; and a
well-defined momentum corresponds to a less well-defined position for the free
particles. In the actual calculations, we have obtained many single-particle
states corresponding to the free particles. Note that to make the RMPD clear, only a part
of free states are displayed in the figure.

\begin{figure}[tbp]
\includegraphics[width=8.5cm]{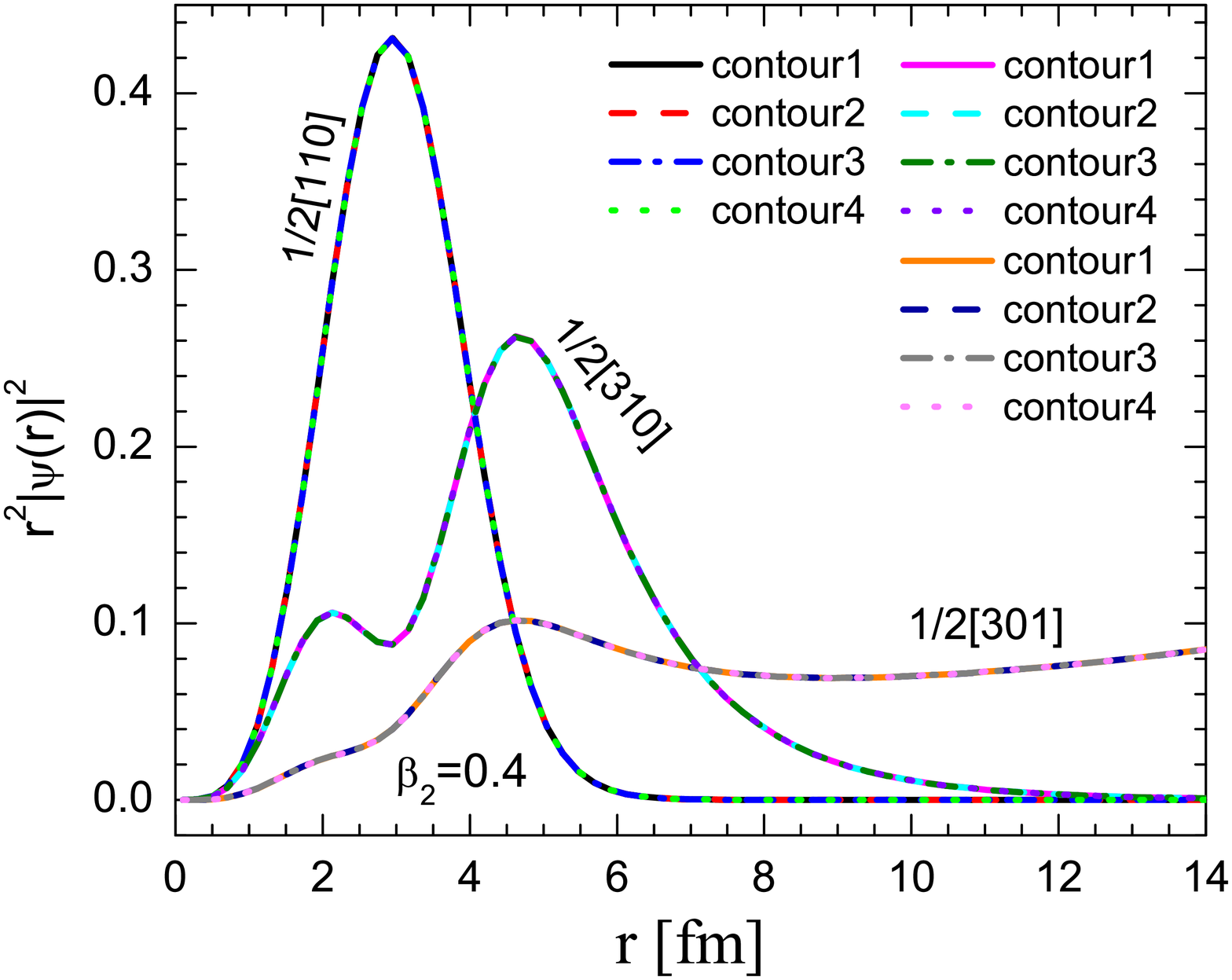}
\caption{(Color online) Radial density distributions in the coordinate space for the bound states 1/2[110] and 1/2[310], and the resonant state 1/2[301] with $\beta_2=0.4$, where the four contours are the same as those in Fig.~1.}
\label{Fig7}
\end{figure}

With these wave functions in the momentum space obtained, we can transform them into the coordinate space by using Eq.~(\ref{radialwavef}). In Fig.~\ref{Fig7}, we have shown the radial density distributions in the coordinate space for the bound states 1/2[110] and 1/2[310], and the resonant state 1/2[301] with $\beta_2=0.4$, where the four contours are the same as those in Fig.~1. Whether bound states or resonant states, the radial density distributions in the coordinate space are independent of the contour, while those of continuous spectra depend on the contour.

The above results indicate that the present method is applicable and
efficient for exploring the resonances in both spherical and deformed
nuclei. Comparing with those frequently used methods that are only
effective for the narrow resonances, the present method is superior because it
is not only appropriate for the narrow resonance, but also can be reliably
applied to the broad resonances that were difficult to obtain before.

\section{Summary}

In summary, we have developed a new method to explore the resonances for
deformed nuclei by solving the Dirac equation in complex momentum representation. In this scheme, the Dirac equation describing deformed nuclei is processed into a set of coupled differential equations by the coupled-channel method. The set of coupled differential equations is then solved by using the complex momentum representation technique, which makes the solutions of Dirac equation become the diagonalization of a matrix. This method describes the bound states and resonant states on the equal footing, which greatly simplifies the problem of how to handle the unbound states for deformed systems.

We have presented the theoretical formalism, elaborated the numerical details, and discussed the dependence of the calculations on the contour of momentum integral, and the satisfactory results are obtained in comparison with the coordinate representation calculations. As an illustrated example, we have explored the resonances in $^{37}$Mg and obtained the energies and widths of single-particle resonant states and their evolutions to the deformation. Compared with the CSM and ACCC calculations, the agreeable results are obtained for narrow resonances. However, for broad resonances that are difficult to be obtained by other methods, the present
method is also applicable and effective.

\section{Acknowledgments}

This work was partly supported by the National Natural Science Foundation of
China under Grant Nos.~11575002, 11175001, 11205004, 11305002, and 11375022; the
Program for New Century Excellent Talents at the University of China under
Grant No. NCET-05-0558; the Natural Science Foundation of Anhui Province
under Grant No. 1408085QA21; the Key Research Foundation of Education
Ministry of Anhui Province of China under Grant No. KJ2016A026; and the 211
Project of Anhui University.


\end{document}